\documentclass[11pt]{article}

\newcommand{\p}{\partial}
\newcommand{\bfT}{\mbox{\bf T}}
\newcommand{\bfnabla}{\mbox{\boldmath $\nabla$}}
\newcommand{\bfe}{\mbox{\bf e}}

\begin{document}
\vspace*{1.in}

\begin{center}
{\Large {\bf Manifestly Covariant Relativity
\footnote{\small This paper appeared in the {\it Hadronic Journal}
{\bf 17}, 139 (1994).  It pre-dates the four related articles
which are listed on page 5.}}}\\
\vspace{.5in}

{\large {\bf Kenneth Dalton}}\\

\vspace{.2in}
e-mail: kxdalton@yahoo.com

\vspace{1.in}

{\bf Abstract}
\end{center}

\vspace{.25in}

According to Einstein's principle of general covariance, all laws of nature
are to be expressed by manifestly covariant equations.  In recent work, the
covariant law of energy-momentum conservation has been established.  Here, we
show that this law gives rise to a fully covariant theory  of gravitation
and that Einstein's field equations yield total energy-momentum conservation.

\clearpage

\begin{quotation}

``The general laws of nature are to be expressed by equations which hold good
for all systems of coordinates, that is, are covariant with respect to any
substitutions whatever (generally covariant).''[1]

\end{quotation}

  The great beauty of Einstein's theory of general relativity has been justly
celebrated over the decades.  The mathematical beauty of the theory consists
in the property of manifest covariance.  The gravitational field equations are
covariant, as are the accompanying equations of motion and Maxwell's field
equations.  The stress-energy-momentum of matter and electromagnetism form
covariant expressions, as well.  Finally, the covariant Bianchi identities
provide closure to the overall geometric structure.  Nevertheless, there comes
a point at which the beauty and simplicity of the theory abruptly end.
In his treatment of energy-momentum, Einstein expressed the differential law
of conservation in terms of the ordinary divergence [2]

\begin{equation}
    \frac{\p \sqrt{-g}\, T^{\mu\nu}}{\p x^\nu} = 0
\end{equation}
This gave rise to the gravitational stress-energy-momentum pseudo\-tensors
[1--6].  These energy pseudo\-tensors are decidedly lacking in simplicity.
They are not covariant expressions, therefore the components can take on
arbitrary values, including zero [3,4].  The conclusion has been reached that
the pseudo\-tensors can have no physical meaning [3], and that no definite 
statement can be made regarding the density of gravitational energy in any
given region of space-time [5].  It is said that gravitational energy is 
`non-localizable', that it is not observable [6], and that it is meaningless
to talk of whether or not there is gravitational energy at a given place [2].
All of this stands in marked contrast to the stress-energy-momentum of matter
and electromagnetism, which is both perfectly well-defined and physically
observable.

Recently, it has been shown that equation (1) does not properly account for
energy-momentum.  The correct expression of conservation is given by the 
covariant divergence [7,8]

\begin{equation}
  T^{\mu\nu}_{;\nu} 
     = \frac{1}{\sqrt{-g}}\frac{\p \sqrt{-g}\, T^{\mu\nu}}{\p x^\nu}
         + \Gamma^{\mu}_{\nu\lambda} T^{\nu\lambda}
     = 0
\end{equation}
This equation arises from the vector divergence formula

\begin{equation}
    \oint {\bfe}_\mu \sqrt{-g}\, T^{\mu\nu}\, d^3 V_\nu
       = \int {\bfe}_\mu \sqrt{-g}\, T^{\mu\nu}_{;\nu}\, d^4 x
\end{equation}
and continuity equation

\begin{equation}
     T^{\mu\nu}_{;\nu} = 0
\end{equation}
It has also been derived by variation of the action

\begin{equation}
    S = \int L \sqrt{-g}\, d^4 x
\end{equation} 
under uniform displacement in space-time [7].  This new law of conservation is
the key to forming a fully covariant theory of gravitation.  We begin with 
Einstein's gravitational field equations

\begin{equation}
   R^{\mu\nu} - \frac{1}{2} g^{\mu\nu} R = - \frac{8\pi G}{c^4} T^{\mu\nu}
\end{equation}
$ T^{\mu\nu} $ represents the stress-energy-momentum of electromagnetism and
matter.  The covariant divergence of the left-hand side is identically zero, 
therefore

\begin{equation}
    T^{\mu\nu}_{;\nu} = 0
\end{equation}
This equation means that the energy-momentum of matter and electromagnetism
is conserved.  In other words, there is no exchange of energy-momentum with
the gravitational field.  The inevitable conclusion to be drawn from this is
that the gravitational field, itself, has no dynamical content---no energy, 
momentum, or stress.  The elimination of the energy pseudotensor removes the 
last vestige of non-covariance from Einstein's theory of gravitation.  All
physically meaningful expressions are now manifestly covariant.  

The absence of gravitational field energy implies that gravitational forces 
do not exist.  We can demonstrate force-free planetary motion by means of a
direct appeal to the equations of motion.  The equation of motion for 
charged matter is found by substituting the energy tensors of matter and 
electromagnetism into equation (7) and then making use of Maxwell's
equations:

\begin{equation}
  \rho c^2 \left(u^\nu \frac{\p u^\mu}{\p x^\nu} 
      + \Gamma^{\mu}_{\nu\lambda} u^\nu u^\lambda \right) 
                       + F^\mu{}_\alpha J^\alpha = 0 
\end{equation}
The path of an uncharged particle satisfies the geodesic equation

\begin{equation}
    \frac{du^\mu}{ds} + \Gamma^{\mu}_{\nu\lambda} u^\nu u^\lambda = 0
\end{equation}
Einstein has stated that the term $ \Gamma^{\mu}_{\nu\lambda} u^\nu u^\lambda $
is to be interpreted as follows: ``The gravitational field transfers energy and
momentum to the matter, in that it exerts forces upon it and gives it energy''
[9].  However, all connection coefficients $ \Gamma^{\mu}_{\nu\lambda} $ can
be transformed to zero in any given infinitesimal region (geodesic coordinates)
[10].  Therefore, no real physical transfer of energy-momentum may be assigned
to this term.  (By way of contrast, the Lorentz force $ F^\mu{}_\alpha J^\alpha$
is covariant and cannot be transformed to zero.)  We conclude that a planet
moves through the gravitational field in force-free geodesic motion.  The field
itself is inherently  curved, and this is revealed by the curved planetary
trajectory.  Only a real physical force, such as the Lorentz force, could
transfer energy-momentum to the planet and cause it to deviate from its 
geodesic path.

The components of the matter tensor $ T^{\mu\nu} = c^2 \rho u^\mu u^\nu $
obviously change during geodesic motion.  The question then arises as to
whether the energy-momentum of the planet is, in fact, conserved.  The
physical density of stress-energy-momentum is not given merely by the above
tensor components, but by the coordinate-independent expression

\begin{equation}
   \bfT = {\bfe}_\mu \otimes {\bfe}_\nu \, T^{\mu\nu}
\end{equation}
In order to investigate conservation, we form the divergence [11]

\begin{equation}
  \bfnabla \cdot \bfT 
     = \bfnabla \cdot \left({\bfe}_\mu \otimes {\bfe}_\nu
        \, T^{\mu\nu}\right)
     = {\bfe}_\mu \, T^{\mu\nu}_{;\nu}
\end{equation}
This is zero by virtue of the equation of motion.  As a planet moves along
its orbit, the basis vectors ${\bfe}_\mu$ change in magnitude and direction
such that $ \bfnabla \cdot \bfT = 0 $ and the energy-momentum of the planet is
conserved.  Similar considerations apply to time-dependent gravitational fields,
in general, and to the emission and absorption of gravitational waves, in
particular.  Conservation of total energy-momentum is assured, if Einstein's
gravitational field equations are satisfied.

\clearpage

\section*{\large {\bf References}}

\begin{enumerate}

\item A. Einstein, ``The Foundation of the General Theory of Relativity''\newline
 in {\it The Principle of Relativity}, (Dover, New York 1952).
\item L.D. Landau and E.M. Lifshitz, {\it The Classical Theory of Fields},
     \newline (Pergamon, 4th edition, 1975) section 96.
\item W. Pauli, {\it The Theory of Relativity}, (Dover, New York, 1981) 
      \newline section 61.
\item H. Weyl, {\it Space-Time-Matter}, (Dover, New York, 1952) section 33.
\item R. Adler, M. Bazin, and M. Schiffer, {\it Introduction to General 
      Relativity}, (McGraw-Hill, New York, 1965) page 314.
\item C. Misner, K. Thorne, and J. Wheeler, {\it Gravitation}, (Freeman,
    \newline  New York, 1973) page 467.
\item K. Dalton, {\it Gen.Rel.Grav.}, {\bf 21}, 533 (1989).
\item J. Vargas and D. Torr, {\it Gen.Rel.Grav.}, {\bf 23}, 713 (1991).
\item A. Einstein, {\it The Meaning of Relativity}, (Princeton, 1953) page 83;
      A.~Einstein (Ref. 1) page 151.
\item L.D. Landau and E.M. Lifshitz (Ref. 2) page 241.
\item C. Misner, K. Thorne, and J. Wheeler, (Ref. 6) page 261.
 
\end{enumerate}
\vspace{.2in}
\section*{\large Related works by the same author (at www.arxiv.org):}

\begin{itemize}

\item ``Farewell to General Relativity''              physics/9710001 
\item ``Einstein's Violation of General Covariance''  physics/9703023 
\item ``Gravity, Geometry, and Equivalence''          gr-qc/9601004 
\item ``Einstein's Energy-Free Gravitational Field''  gr-qc/9512008 

\end{itemize}

\end{document}